\documentclass[a4paper,10pt]{iopart}
\usepackage[utf8x]{inputenc}
\usepackage{graphicx}   
\newcommand{\figpath}{.}

\usepackage{amscd}
\usepackage{iopams}

\newcommand{\ri}{{ \rm i }}

\newcommand{\rd}{{ \rm d }}
\newcommand{\rR}{{ \rm R }}
\newcommand{\rL}{{ \rm L }}

\newcommand{\be}{\begin{equation}}
\newcommand{\ee}{\end{equation}}

\usepackage{color}
\definecolor{blau}{rgb}{0,0,1}
\definecolor{gruen}{rgb}{0,1,0}
\definecolor{rot}{rgb}{1,0,0}
\definecolor{magenta}{rgb}{1,0,1}

\begin{document}

\title[Heuristic approach to BEC self-trapping in double wells beyond mean-field]{Heuristic approach to BEC self-trapping in double wells beyond mean-field}

\author{K. Rapedius$^{1,2}$} 
\address{$^1$ FB Physik, TU Kaiserslautern, D-67653 Kaiserslautern, Germany}
\address{$^2$ Center for Nonlinear Phenomena and Complex Systems, Universit\'e Libre de Bruxelles (ULB), Code Postal 231, Campus Plaine,
B-1050 Brussels, Belgium}

\ead{rapedius@physik.uni-kl.de}

\begin{abstract}
We present a technically simple treatment of self-trapping of Bose-Einstein condensates in double well traps based on intuitive semiclassical approximations. 
Our analysis finally leads to a convenient closed form approximation for the time-averaged population imbalance valid in both the mean-field case and
in the case of finite particle numbers for short times.
\end{abstract}

\date{\today}



\maketitle

\section{Introduction}

Despite its simplicity an atomic Bose-Einstein condensate (BEC) in a double well trap is a quantum system showing a rich dynamical behaviour in different parameter ranges that
can be quite accurately controlled in current experiments (see e.~g.~\cite{Albi05,Mors06,Gati07} and references therein). Even in the mean-field limit of high particle numbers $N$ where many-particle 
correlations can be neglected, different dynamical regimes are observed. The latter can be understood by mapping the system to a classical pendulum the dynamics of which can be analysed in the corresponding two-dimensional 
phase space (see e.g.~\cite{Ragh99}). For small and moderate interaction $U$ between the particles a condensate that is initially localized in one of the two wells performs Rabi respectively Josephson oscillations between the two wells. 
If the inter-particle interaction exceeds a critical value, the initially occupied well remains macroscopically occupied due to spontaneous symmetry breaking. 
This is referred to as running phase self-trapping effect.

While for finite particle numbers the self-trapping effect is eventually destroyed by quantum correlations in the long-time limit, it can still be observed for a considerable time span depending on 
the number of particles in the system \cite{Kalo03,Kalo03a}. In this ``short time regime`` the main dynamical features of the system are well described by (semi-) classical phase space methods
like the truncated Wigner approximation \cite{Chuc10} or a similar ansatz based on the Husimi distribution \cite{07phase,09phaseappl} where the respective quantum-mechanical phase space representations of an 
initial quantum state are sampled by an ensemble of trajectories that are propagated according to the corresponding Gross-Pitaevskii mean-field equation. 
Related aspects of the system studied in the literature include the semiclassical WKB-type quantization of its energy spectrum \cite{07semiMP} and the quantum fluctuations of the eigenstates near the point of transition to 
self-trapping \cite{Juli10}. 

In this paper, following the example of \cite{Fu06,Liu07}, the transition to self-trapping is described by means of a single scalar quantity, namely the time-average $\bar{z}(\Lambda)$ of the relative population 
imbalance between the two wells as a function of the scaled interaction parameter $\Lambda=U(N-1)/J$ where $J$ is the tunneling coefficient (cf.~equation (\ref{Ham}) below).
It will be shown that a useful closed form approximation for $\bar{z}(\Lambda)$ in the aforementioned semiclassical parameter and time regimes can be derived
in a technically simple manner by means of intuitive heuristic approximations. 

This article is organized as follows: In section \ref{sec:Ham} we discuss some selected properties of the system Hamiltonian, namely the Bose-Hubbard dimer, and its mean-field limit. 
In section \ref{sec:mf} we use a heuristic ansatz to derive a closed form approximation for the time-averaged population imbalance as a 
function of the scaled interaction strength for the mean-field case. In section \ref{sec:fluct} many-particle effects are taken into account 
in a semiclassical manner by incorporating quantum fluctuations of the initial state into the mean-field description derived in the 
previous section which again leads to a closed form approximation for the time-averaged population imbalance. 
The main results are briefly summarized in section \ref{sec:summary}. \ref{sec:app} contains a brief review of the dynamics of the corresponding noninteracting system.

\section{The Hamiltonian}
\label{sec:Ham}
In the two mode approximation the double well system is described by the Bose-Hubbard dimer Hamiltonian (cf.~e.g. \cite{Gati07})
\begin{equation}
\fl \quad \quad  {\hat H}= - \frac{J}{2} \left({\hat a}^\dagger_\rL {\hat a}_\rR +{\hat a}^\dagger_\rR {\hat a}_\rL   \right) + \frac{U}{2} \left( ({\hat a}^\dagger_\rL)^2 {\hat a}_\rL^2 + ({\hat a}^\dagger_\rR)^2 {\hat a}_\rR^2 \right)+ \epsilon_\rL {\hat a}^\dagger_\rL {\hat a}_\rL + \epsilon_\rR {\hat a}^\dagger_\rR {\hat a}_\rR 
\label{Ham}
\end{equation}
where ${\hat a}_\rL$ and ${\hat a}_\rR$ are the annihilation operators of a particle in the left and right well respectively. 
$J>0$ determines the rate of tunneling between adjacent lattice sites, $U$ is the inter particle interaction parameter and $\epsilon_\rL$, $\epsilon_\rR$ are the on-site energy terms. 
Here, we concentrate on repulsive interaction $U>0$. 


For a sufficiently high number of particles $N$ in the system, one can apply a Hartree-Fock mean-field approximation (see e.g.~\cite{Cast00}) where the operators ${\hat a}_{L,R}$ in the Hamiltonian 
can be replaced by the complex numbers $\sqrt{N} c_{L,R}$ representing their respective coherent state expectation values. This leads to the ``classical`` Hamiltonian
\begin{equation}
\fl \quad \quad  H/N= - \frac{J}{2} \left( c_\rL^* c_\rR + c_\rR^* c_\rL \right) + \frac{U}{2} (N-1) \left( |c_\rL|^4 +|c_\rR|^4 \right) +\epsilon_\rL |c_\rL|^2 +\epsilon_\rR |c_\rR|^2 \,. \quad \quad
\label{Ham_class}
\end{equation}
The canonical equations of motion $i \hbar \dot c_l =\partial H / \partial c_l^*$, $i \hbar \dot c_l^* =-\partial H / \partial c_l$ then lead to the coupled discrete Gross-Pitaevskii equations 
or nonlinear Schr\"odinger equations
\begin{eqnarray}
  i \hbar \dot c_\rL=- \frac{J}{2}c_\rR +\left(\epsilon_\rL + U(N-1) |c_\rL|^2 \right) c_\rL  \label{GPE1}\\
  i \hbar \dot c_\rR=- \frac{J}{2}c_\rL + \left(\epsilon_\rR + U(N-1) |c_\rR|^2 \right) c_\rR  \,. \label{GPE2} 
\end{eqnarray}
which describe the mean-field dynamics of the on-site amplitudes. Particle number conservation yields $|c_\rL|^2+|c_\rR|^2=1$.

Introducing the amplitude-phase representation $c_\rR=\sqrt{p}\exp(\ri \theta_\rR)$, $c_\rL=\sqrt{1-p}\exp(\ri \theta_\rL)$ the classical Hamiltonian (\ref{Ham_class}) reads
\begin{equation}
\fl \quad \quad   H/N= - J \sqrt{p(1-p)} \cos \theta + \frac{U}{2} (N-1) \left((1-p)^2+p^2 \right)+ \epsilon_\rL (1-p) + \epsilon_\rR p 
\end{equation}
with $\theta=\theta_\rR-\theta_\rL$.

In this paper we consider the situation of a symmetric double well where $\epsilon_\rL=0=\epsilon_\rR$ and initial conditions where all particles occupy the left site, 
i.e.~$|c_\rL|^2=1$, $|c_\rR|^2=0$ respectively $p=0$ at time $t=0$. Then the corresponding conserved energy per particle reads $E/N=H(p=0)/N=U(N-1)/2$.
We assume that self-trapping occurs if an equal population of both sites, i.e~ $|c_\rL|^2=1/2=|c_\rR|^2$ respectively $p=1/2$ is no longer possible. Conservation of energy yields 
(cf.~ e.g.~\cite{Ragh99}) $E/N=H(p=1/2)/N$ which leads to the condition $\cos \theta=U(N-1)/(2J)$. $|\cos \theta| \le 1$ implies that an 
equal occupation of both sites is only possible for $|U(N-1)/J| \le 2$ so that self-trapping occurs if
\be
   \Lambda \equiv \frac{U(N-1)}{J}  > 2 
\ee
(for $U>0$, $J>0$). 
Please note that the effect considered here is referred to as running phase self-trapping and should not be confused with
so-called $\pi$-phase self-trapping that is closely related to the appearance of symmetry breaking nonlinear eigenstates of the 
Gross-Pitaevskii equations (\ref{GPE1}), (\ref{GPE2}) (see e.g.~\cite{Ragh99,10nhbh} for a more detailed discussion).

\section{Heuristic mean-field approach}
\label{sec:mf}

In the mean-field case, described by the Hamiltonian (\ref{Ham_class}) and the corresponding Gross-Pitaevskii equations (\ref{GPE1}), 
(\ref{GPE2}), the site populations $|c_\rL|^2$ and $|c_\rR|^2$ are periodic in time \cite{Ragh99}.
In the following we can thus quantify the self-trapping by means of the time-averaged relative population imbalance (cf.~\cite{Fu06,Liu07})
\be
  \bar{z} = \overline{|c_\rL|^2-|c_\rR|^2}= 1-2\bar{p} \, \quad -1 \le \bar{z} \le 1 \label{zbar_def}
\ee
where the overbar denotes time averaged quantities and $p=|c_\rR|^2$ as introduced in the previous section.
For the mean-field Gross-Pitaevskii case an exact solution in terms of elliptic integrals was derived in \cite{Fu06}. 
Here we instead aim at a convenient approximation in terms of elementary functions.

To this end we first consider the noninteracting limit of our system, as described by equations (\ref{GPE1}), (\ref{GPE2}) with $U=0$.
If at time $t=0$ all particles are in the left well, the occupation $|c_\rR(t)|^2=p(t)$ in the noninteracting two state quantum system is given by (see e.g.~\cite{Cohe99})
\be
   p(t)=\frac{J^2}{J^2+\Delta^2} \sin^2\left( \frac{\sqrt{J^2+\Delta^2}}{2 \hbar}t \right) 
\label{pt_lin}
\ee
where $\Delta=\epsilon_\rL-\epsilon_\rR$ is the difference of the on-site chemical potentials. For convenience a brief derivation of this standard result is given in \ref{sec:app}. 

In the interacting case with a finite $U>0$ the time-dependent local chemical potentials of the two sites read $\epsilon_\rL+U(N-1)|c_\rL(t)|^2$ 
and $\epsilon_\rR+U(N-1)|c_\rR(t)|^2$ respectively (cf.~(\ref{GPE1}), (\ref{GPE2})). We therefore introduce the interaction into (\ref{pt_lin}) 
in a heuristic manner by replacing $\Delta$ with the time-averaged difference of the on-site terms,
i.e.~we set $\Delta=\epsilon_\rL-\epsilon_\rR+U(N-1)(\overline{|c_\rL|^2-|c_\rR|^2})=\epsilon_\rL-\epsilon_\rR+J \Lambda (1-2\bar{p})$. 
For the special case of a symmetric double well where $\epsilon_\rL=0=\epsilon_\rR$ equation (\ref{pt_lin}) thus becomes
\be
   p(t) = \frac{1}{1+\Lambda^2(1-2\bar{p})^2} \sin^2\left( \sqrt{1+\Lambda^2(1-2\bar{p})^2}\frac{J}{2 \hbar}t \right) \,.
\label{pt}
\ee
Self-consistency then requires
\be
   \bar{p} = \frac{1}{2} \frac{1}{1+\Lambda^2(1-2\bar{p})^2}
\label{pbar}
\ee
where we have inserted the time average $\overline{\sin^2t}=1/2$. It is convenient to replace $\bar{p}$ by $\bar{z}$ via (\ref{zbar_def}), 
\be
    \bar{z} = 1- \frac{1}{1+\Lambda^2 \bar{z}^2} \,,
\label{zbar_self}
\ee
which yields the cubic equation
\be
   \bar{z}^3 -\bar{z}^2 +\frac{1}{\Lambda^2}\bar{z}=0 \,.
\label{zbar_cubic}
\ee
For $\Lambda<2$ equation (\ref{zbar_cubic}) has only one solution $\bar{z}=0$. As shown in the previous section, conservation of energy requires that $\bar{z}>0$ (i.e.~population in the left well higher than in the right one)
if $\Lambda>2$. Thus we obtain
\be
   \bar{z}(\Lambda)=\left\{
                \begin{array}{cl}
        	0 \,, & |\Lambda| \le 2\\
		\frac{1}{2}+\sqrt{\frac{1}{4}-\frac{1}{\Lambda^2}}\,,& |\Lambda| >2
		\end{array}
        \right.
\label{zbar}
\ee
for the time-averaged relative population imbalance. The negative square root solution $\bar{z}(\Lambda)=1/2-(1/4-1/\Lambda^2)^{1/2}, \, |\Lambda| >2$
is discarded since it does not yield the correct limit $\bar{z} \rightarrow 1$ for $\Lambda \rightarrow \infty$.

\begin{figure}[htb]
\begin{center}
\includegraphics[width=0.5\textwidth] {\figpath/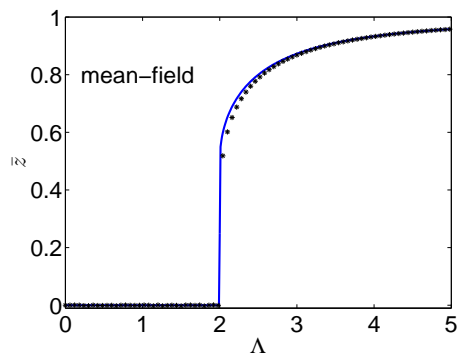}
\caption{\label{fig:mean-field} Numerically exact time-averaged relative population imbalance $\bar{z}$ (*) as a function of the scaled interaction strength $\Lambda=U(N-1)/J$ compared to the approximation (\ref{zbar}) (\textcolor{blau}{-}). } 
\end{center}
\end{figure}
Figure \ref{fig:mean-field} compares the approximation (\ref{zbar}) with the numerically exact values obtained by integrating the Gross-Pitaevskii equations (\ref{GPE1}), (\ref{GPE2}). Considering the simplicity of our approach we observe a good agreement between the curves. 
The deviations for $\Lambda \gtrsim 2$ are mainly due to the fact that the ansatz does not take into account the change in the shape of the oscillations from sinusoidal 
to Jacobi elliptic caused by the nonlinearity (cf.~\cite{Ragh99}).

\section{Heuristic semiclassical ansatz: Adding quantum fluctuations}
\label{sec:fluct}
\begin{figure}[htb]
\begin{center}
\includegraphics[width=0.45\textwidth] {\figpath/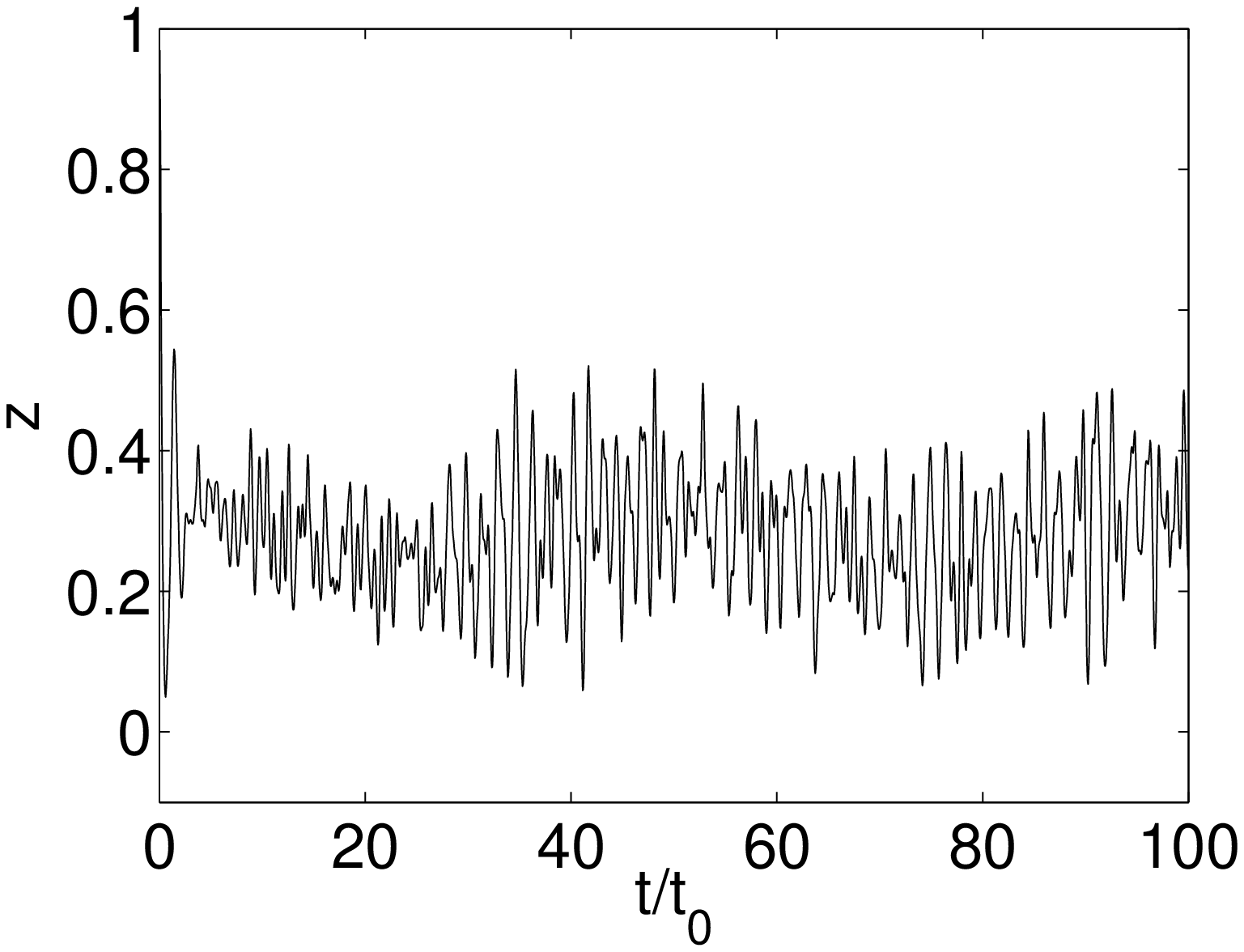}
\includegraphics[width=0.45\textwidth] {\figpath/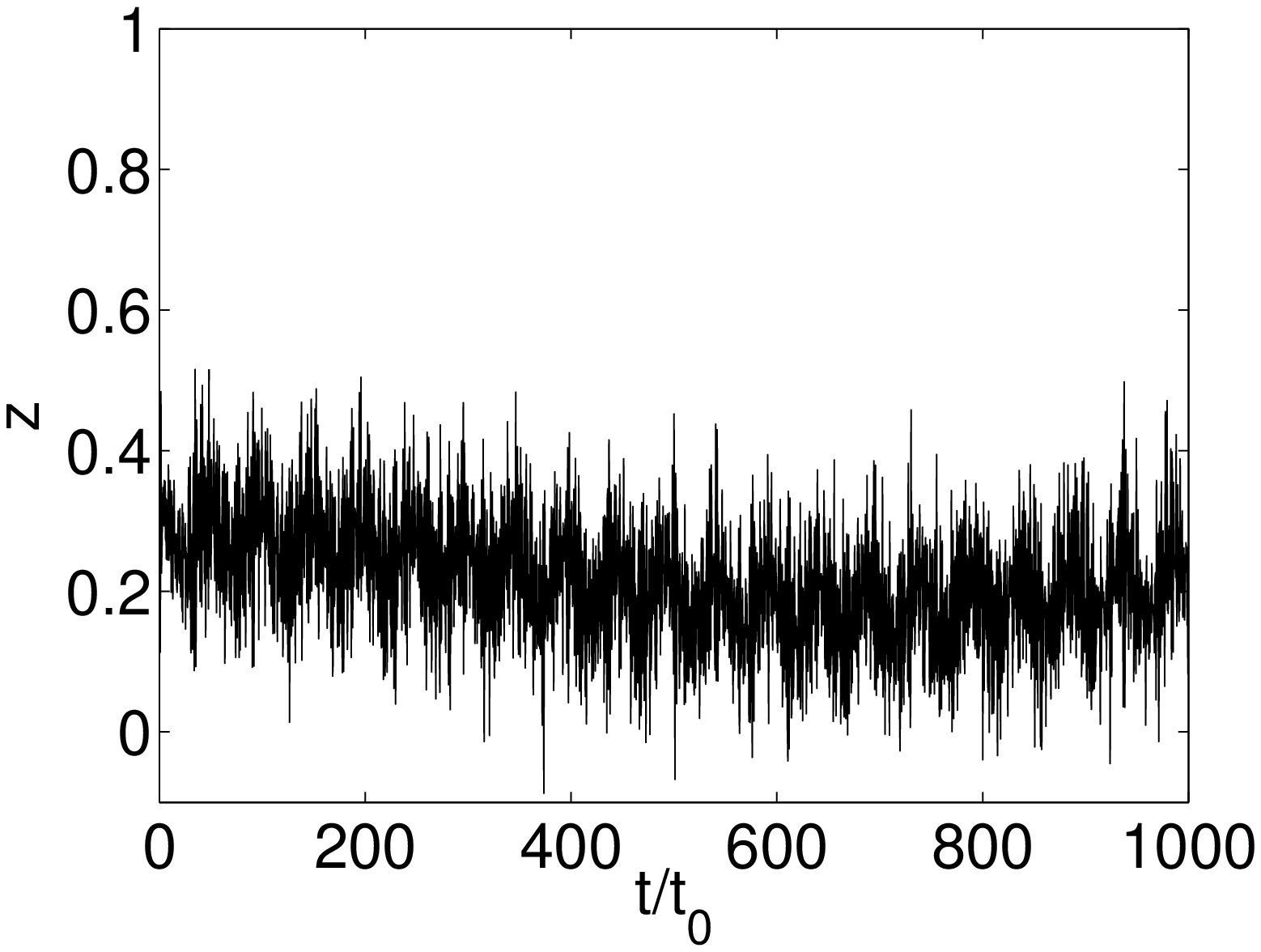}
\caption{\label{fig:single_traj} Numerically exact population imbalance $z(t)$ as a function of time for $N=100$, 
$\Lambda=2$ and $J=1$. The time is given in multiples of $t_0=2 \pi \hbar/J$. } 
\end{center}
\end{figure}

In this section we consider the Bose-Hubbard dimer (\ref{Ham}) with a finite number $N$ of particles. In \cite{Kalo03,Kalo03a} the dynamics of the initial
state $|N,0 \rangle$, i.~e. the number state where all particles are in the left well, was analysed in the regime of strong
interaction by treating the tunneling terms proportional to $J$ as a small perturbation. 
It was found that the self-trapping effect is destroyed by quantum correlations in the long-time limit but that it can still be observed for 
a considerable time span that quickly increases with the number of particles in the system. Even for moderate interaction strengths, 
where the perturbative treatment of the tunneling terms is not a good approximation, the self-trapping effect can still be observed for a 
considerable time span before it is destroyed. This is illustrated in figure \ref{fig:single_traj} which shows the time-dependent 
population imbalance $z(t)$ for $N=100$ and $\Lambda=2$ for the initial state $|N,0 \rangle$. The time is given in multiples of $t_0=2 \pi/\omega_0$ where
$\omega_0=J/\hbar$ is the Rabi frequency in the noninteracting case $U=0$ (cf.~ \ref{sec:app}).
For shorter times the system shows self-trapping similar to the mean-field case (left panel). For longer times an oscillation
with a higher amplitude and a longer period is revealed (right panel). 
For even longer times, which are beyond the range of numerical 
accuracy for the given parameters, the self-trapping effect is expected to be completely destroyed with the system showing a total 
population transfer between the two sites (cf.~discussion in \cite{Kalo03,Kalo03a}).

In the short time regime the dynamics of the system is quite well described by semiclassical phase space methods where quantum 
mechanical phase representations of the initial quantum state like the Wigner \cite{Chuc10} or Husimi \cite{07phase,09phaseappl} distributions are propagated in 
phase space according to the Gross-Pitaevskii equations (\ref{GPE1}), (\ref{GPE2}).

In the spirit of these semiclassical methods we want to add the quantum fluctuations of the initial state to the simple 
mean-field description developed in the last section. Since the expression (\ref{zbar_self}) for the time-averaged relative population 
imbalance only depends explicitly on the amplitude of the oscillation (and thus on the local particle numbers in the wells) but not on the relative phase 
between the wells we want to incorporate the quantum fluctuations of the initial state via the fluctuations of the local particle 
numbers. To estimate the particle number fluctuations we consider the approximate mean-field dynamics of the relative particle 
number in the right well given in (\ref{pt}), which can be conveniently rewritten as
\be
   p(t)=p_{\rm amp} \sin^2 (\omega_\Lambda t)
\label{pt2}
\ee 
with the abbreviations $p_{\rm amp}=(1+\Lambda^2(1-2\bar{p})^2)^{-1}$ and $\omega_\Lambda=\sqrt{1+\Lambda^2(1-2\bar{p})^2} J/(2 \hbar)$.
For interaction strengths $\Lambda \gtrsim 2$ near the transition point, where we expect the quantum fluctuations to have their greatest influence, the model 
(\ref{zbar}) predicts $\bar{z} \approx 1/2$ which implies $\bar{p} \approx 1/4$ and $p_{\rm amp} \approx 1/2$ , i.e.~half the particles take part 
in the oscillation. 
\begin{figure}[htb]
\begin{center}
\includegraphics[width=0.45\textwidth] {\figpath/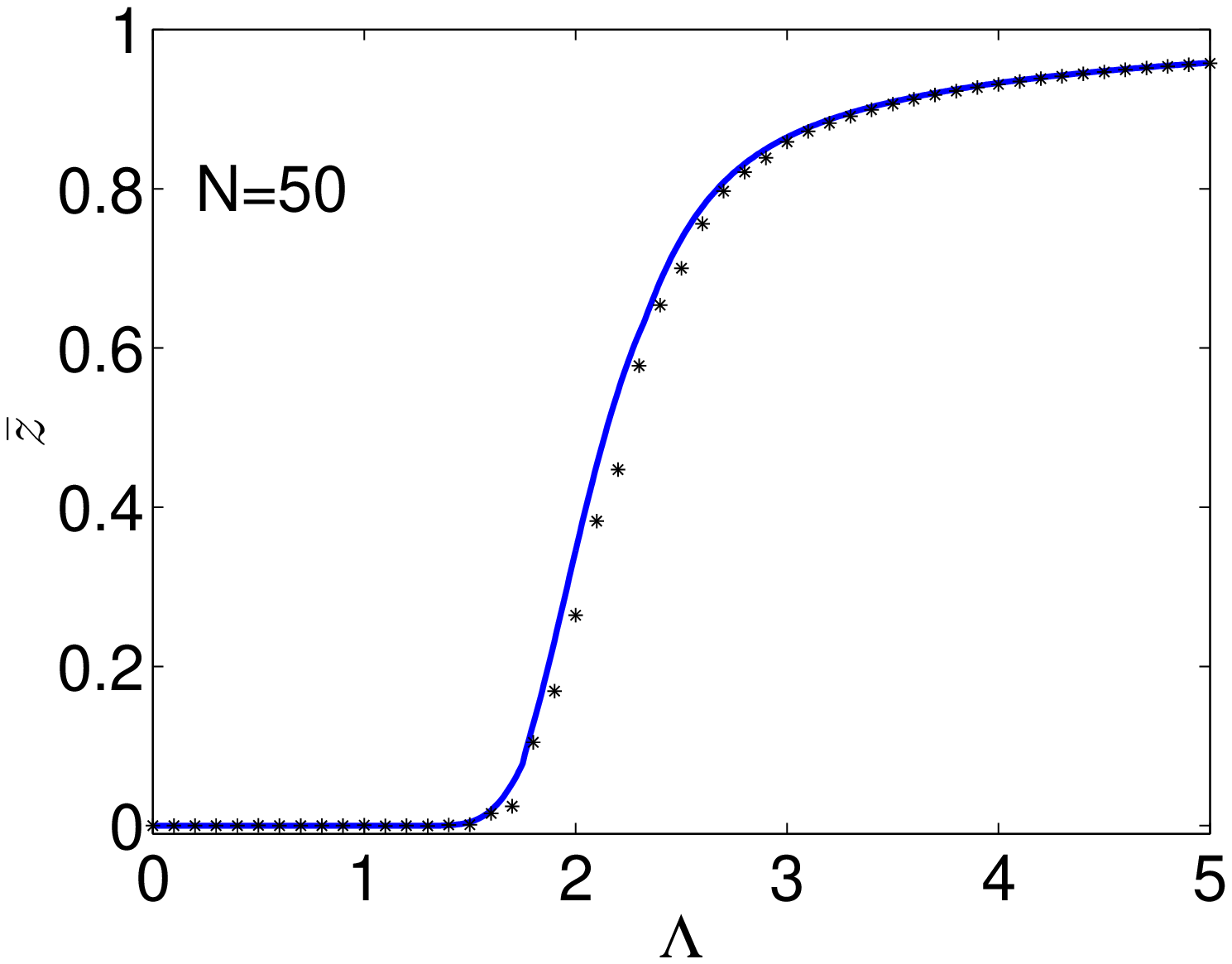}
\includegraphics[width=0.45\textwidth] {\figpath/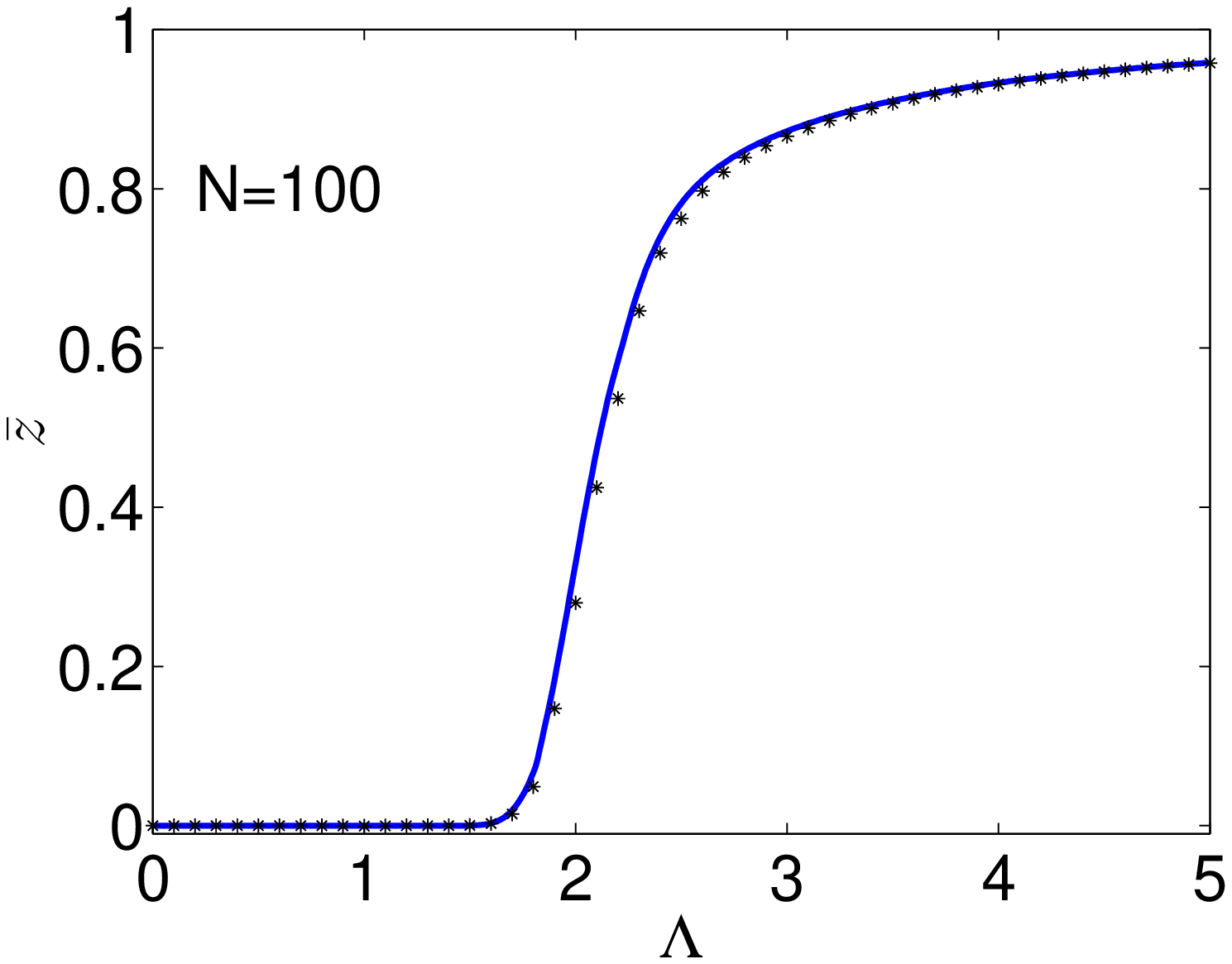}\\
\includegraphics[width=0.45\textwidth] {\figpath/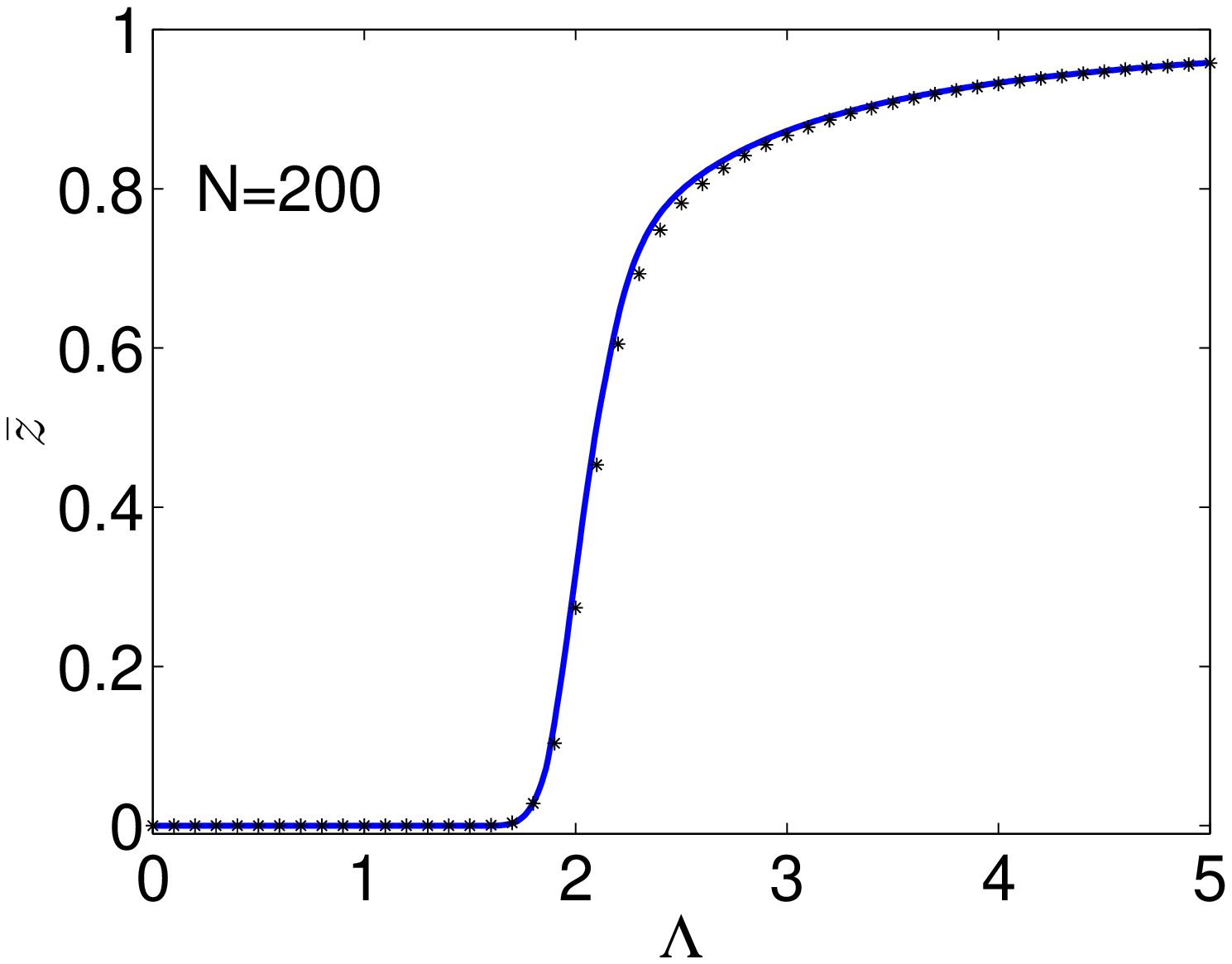}
\includegraphics[width=0.45\textwidth] {\figpath/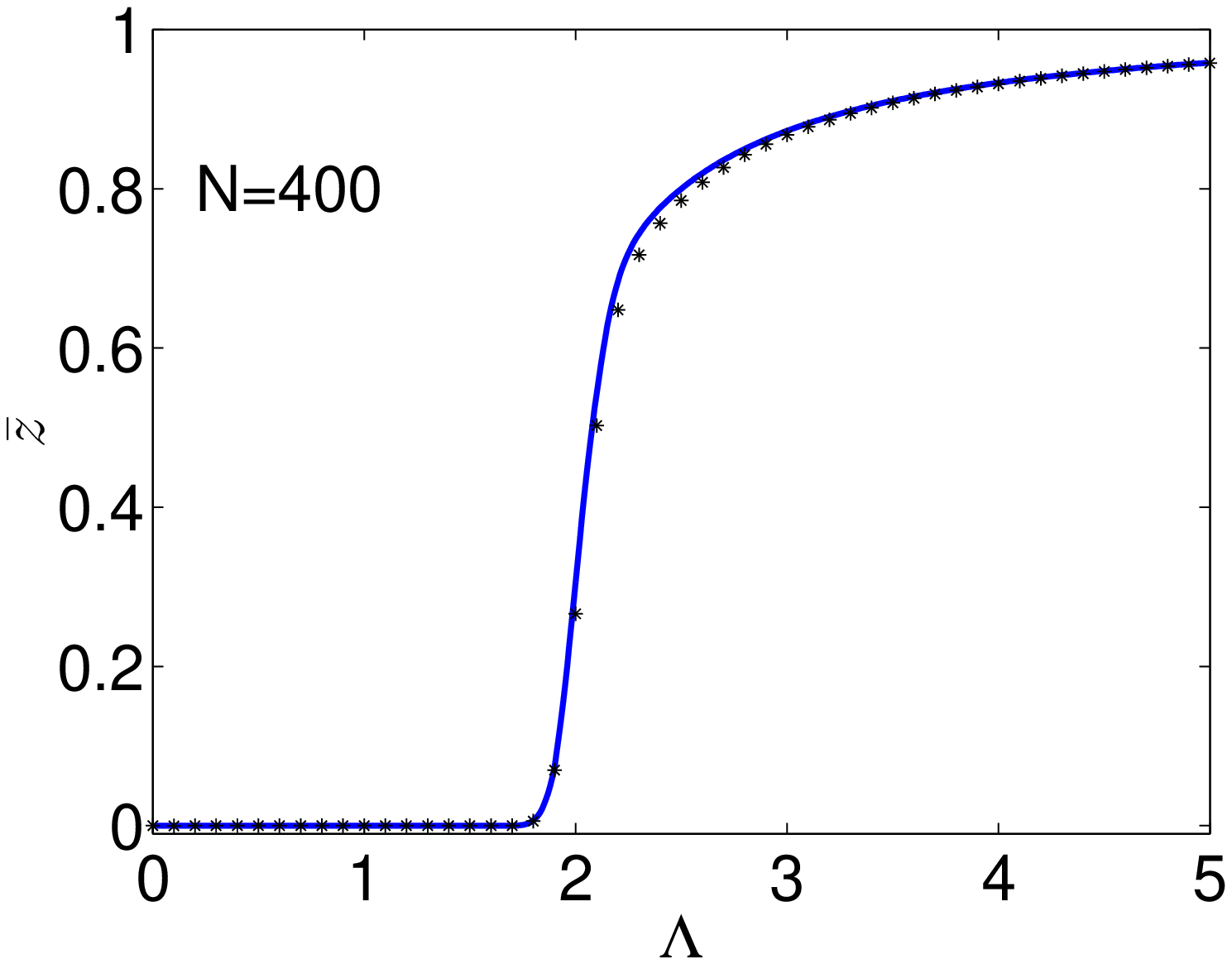}
\caption{\label{fig:N} Numerically exact time-averaged relative population imbalance $\bar{z}$ (*) as a function of the scaled interaction $\Lambda=U(N-1)/J$ compared to the approximation (\ref{zbar_app}) (\textcolor{blau}{-}) 
 for different particle numbers $N$. } 
\end{center}
\end{figure}

Particle number fluctuations of mean-field like states were recently analysed in \cite{Sakm11}. Following \cite{Sakm11}
we define the operators 
\be
  {\hat a}_0=\sqrt{1-p} \, {\hat a}_\rL+ \sqrt{p} \, {\hat a}_\rR \,, \qquad {\hat a}_\perp=-\sqrt{p} \, {\hat a}_\rL+ \sqrt{1-p} \, {\hat a}_\rR 
  \label{a0_ap}
\ee
where ${\hat a}_0$ destroys a state with occupations $1-p$ in the left and $p$ in the right well respectively and
${\hat a}_\perp$ destroys a state orthogonal to the first one. A mean-field like state with $N$ particles and 
occupations $1-p$ in the left and $p$ in the right well is thus given by
\be
  |\psi_0 \rangle = (N!)^{-1/2} ({\hat a}^\dagger_0)^N |0 \rangle \,.
\ee
The inversion of relation (\ref{a0_ap}) reads
\be
  {\hat a}_\rR=\sqrt{p} \, {\hat a}_0+ \sqrt{1-p}  \, {\hat a}_\perp \,, \qquad {\hat a}_\rL=\sqrt{1-p} \, {\hat a}_0- \sqrt{p} \, {\hat a}_\perp \,.
  \label{aR_aL}
\ee
The fluctuations of the particle number ${\hat n}_\rR={\hat a}_\rR^\dagger {\hat a}_\rR$ in the right well are given by $\Delta {\hat n}_\rR^2 = \langle \psi_0|{\hat n}_\rR^2|\psi_0\rangle - \langle \psi_0|{\hat n}_\rR|\psi_0\rangle ^2$. 
Using (\ref{aR_aL}) ${\hat n}_\rR$ can be expressed in terms of ${\hat a}_0$ and ${\hat a}_\perp$ so that the expectation values with respect
to $|\psi_0 \rangle$ can be evaluated in a straightforward manner which yields (cf.~\cite{Sakm11}) 
\be 
  \Delta {\hat n}_\rR^2=N \, p(1-p) \,.
\label{flukt}
\ee 
For an initial state with $p=0$ (i.e.~all particles in the left well) the particle number fluctuations vanish and there are only phase fluctuations which cannot be incorporated 
easily into our classical model (\ref{zbar}) whereas it is quite straightforward, as shall be shown in the following, to include number fluctuations. 
Since the approximate mean-field dynamics (\ref{pt2}) is periodic (as is the exact one, see e.g.~\cite{Ragh99}) each mean-field 
state at some time $t \in [0, \pi/\omega_\Lambda]$ creates the same dynamics (apart from an irrelevant phase shift) if chosen as the initial state. We therefore consider the 
state at $t=\pi/(2 \omega_\Lambda)$ as initial state where $p=p_{\rm amp}$ and particularly $p_{\rm amp}=1/2$ for $\Lambda=2$. For this state the particle number fluctuations 
(\ref{flukt}) assume their maximum $\Delta {\hat n}_\rR^2=N \, p_{\rm amp}(1-p_{\rm amp})=N/4$. This corresponds to a standard deviation of the amplitude 
$p_{\rm amp}=\langle {\hat n}_\rR \rangle/N$ of ${\rm std}(p_{\rm amp})=\Delta {\hat n}_\rR/N=1/(2 \sqrt{N})$. The standard deviation of the time averaged population 
imbalance $\bar{z}$ is then given by 
\be
   {\rm std}(\bar{z})=2 {\rm std}(p_{\rm amp}) \overline{\sin^2(\omega_\Lambda t)}=\frac{1}{2 \sqrt{N}} \,.
\label{stdz}
\ee
Applying the central limit theorem we assume the fluctuations to be Gaussian for the particle numbers $N \ge 50$ considered in the following.

Now the initial quantum fluctuations can be added to our classical description by means of the replacement 
$\bar{z} \rightarrow \bar{z}+\delta \bar{z}$ on the right hand side of (\ref{zbar_self}),
\be
    \bar{z} = 1- \frac{1}{1+\Lambda^2 (\bar{z}+\delta \bar{z})^2} \,,
\label{zbar_self_flukt}
\ee
where $\delta \bar{z}$ is a Gaussian random variable with zero mean and standard deviation (\ref{stdz}).
To lowest order in $\delta \bar{z}$ we obtain the cubic equation
\be
  \bar{z}^3+(2 \delta \bar{z}-1)\bar{z}^2+\left(\frac{1}{\Lambda^2}-2 \delta \bar{z} \right) \bar{z}=0
\ee
which yields, in analogy to (\ref{zbar}),
\be
  \fl \qquad \bar{z}(\Lambda,\delta \bar{z})=\left\{
                \begin{array}{cl}
		0 \,, & (\delta \bar{z}+ \frac{1}{2})^2 -\frac{1}{\Lambda^2}\le 0\\
		\frac{1}{2}-\delta \bar{z}+\sqrt{(\delta \bar{z}+ \frac{1}{2})^2 -\frac{1}{\Lambda^2}}\,,& (\delta \bar{z}+ \frac{1}{2})^2 -\frac{1}{\Lambda^2}>0
		\end{array}
        \right. \,.
\label{zbar_flukt}
\ee
The semiclassical result for the time-averaged relative population imbalance can thus be directly obtained by averaging over (\ref{zbar_flukt}) with $\delta \bar{z}=\zeta/(2 \sqrt{N})$ where
$\zeta$ is a standard Gaussian random variable with ${\rm mean}(\zeta)=0$ and ${\rm std}(\zeta)=1$.

Neglecting fluctuations with $\delta \bar{z}< -1/\Lambda-1/2$ the time-averaged relative population imbalance can alternatively be expressed as the integral
\begin{equation}
 \fl \quad  \bar{z}(\Lambda) = \int_{x_0}^\infty \left(\frac{1}{2}-x+\sqrt{\left(\frac{1}{2}+x \right)^2-\frac{1}{\Lambda^2}}\right)\frac{\exp \left( - \frac{x^2}{2 \sigma_N^2} \right)}{\sqrt{2 \pi \sigma_n^2}} \,  \rd x \\
\end{equation}       
\begin{equation}
\fl \quad \qquad   = \frac{1}{2} \Phi\left(-\frac{x_0}{\sigma_N} \right)- \sigma_N \phi \left(\frac{x_0}{\sigma_N} \right) + \int_{x_0}^\infty \rd x \, \sqrt{\left(\frac{1}{2}+x \right)^2-\frac{1}{\Lambda^2}} \frac{1}{\sigma_N} \phi \left(\frac{x}{\sigma_N} \right) \label{zbar_int}
\label{dings}
\end{equation}
where $x_0=1/\Lambda-1/2$, $\sigma_N=1/(2 \sqrt{N})$, $\phi(x)=\exp(-x^2/2)/\sqrt{2 \pi}$ is the standard normal distribution and
 $\Phi(x)=\int_{-\infty}^{x} \rd x\, \phi(x)=[1+{\rm erf}(x/\sqrt{2})]/2$ the corresponding 
cumulative distribution function. The second term in (\ref{dings}) is small compared to the first and the third one and can be neglected. 
Approximating the square root term under the integral by its value at $x^*={\rm max}(x_0,(x_0+\sigma_N)/2,0)$ we thus arrive at
\begin{equation}
 \bar{z}(\Lambda)  =  \left(\frac{1}{2}+\sqrt{\left(\frac{1}{2}+x^* \right)^2-\frac{1}{\Lambda^2}} \right)\Phi\left(-\frac{x_0}{\sigma_N} \right) \,.
 \label{zbar_app}
\end{equation}

In figure \ref{fig:N} we compare our approximation for $\bar{z}$ (\ref{zbar_app}) with the numerically exact 
results for different values of the particle number $N$. To obtain the numerically exact short time results 
we compute the dynamics of the system in the Fock basis of states $|N-n,n \rangle$, $0 \le n \le N$ for the initial state $|N,0 \rangle$
and average over the time interval $[0, 100 t_0]$. Generally we observe a good agreement between the approximation and the numerical 
results. The small deviations around $\Lambda \approx 2$ that are inherited from the heuristic mean-field approach are related to the 
influence of the mean-field interaction on the shape of the oscillations.
The deviation is less pronounced for smaller particle numbers since the quantum fluctuations lead to deviations from the 
characteristic shape of the mean-field oscillations (cf.~figure \ref{fig:single_traj}) that are given by Jacobi elliptic functions \cite{Ragh99}.  
The additional deviations observed for $N=50$ are caused by the fact that for this relatively small particle number some effects 
of low frequency modes (cf.~figure \ref{fig:single_traj} and the discussion at the beginning of the section) can be felt even for 
short times for interaction strengths around $\Lambda \approx 2$.
Qualitatively, our results confirm the behaviour found in a previous numerical study \cite{Fu06}: The quantum fluctuations cause a broadening 
and softening of the transition region between the Josephson oscillation regime and the self-trapping regime. Quantitatively, however, 
there are some deviations because in \cite{Fu06} the time averages are performed over much longer time intervals such that the system
dynamics is influenced by the low frequency modes mentioned above.

\begin{figure}[htb]
\begin{center}
\includegraphics[width=0.6\textwidth] {\figpath/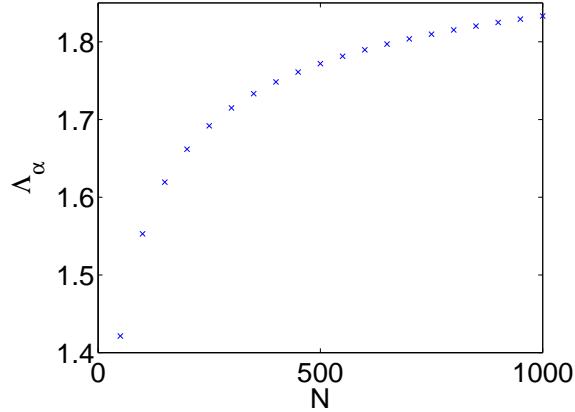}
\caption{\label{fig:Lambda_N} Critical interaction strength (\ref{Lambda_crit}) as a function of the particle number $N$ for
the threshold value $\alpha=0.001$. } 
\end{center}
\end{figure}

Since there is no longer a sharp transition point for finite particle numbers $N$ we quantify the shift of the transition point by
considering the interaction strength $\Lambda$ for which the population imbalance $\bar{z}$ surpasses a certain small threshold value 
$\alpha$ (cf.~\cite{Fu06}). This happens in the region where $x^*=x_0=1/\Lambda-1/2$, so that 
$\bar{z}(\Lambda)  \approx  \Phi\left((1-2/\Lambda)\sqrt{N} \right)/2$. 
Setting this expression equal to $\alpha$ we obtain the condition
\be
   \Lambda_{\alpha} \approx 2 \left(1- N^{-1/2}\Phi^{-1}(2 \alpha) \right)^{-1}
\label{Lambda_crit}
\ee
with the inverse function $\Phi^{-1}(x)=\sqrt{2}{\rm erf}^{-1}(2x-1)$ known as quantile function. For high particle numbers $N \gg 1$
the critical interaction strength behaves like $\Lambda_\alpha \approx 2+2 \Phi^{-1}(2 \alpha) N^{-1/2}$. In the limit $N \rightarrow \infty$ 
one recovers the mean-field result $\Lambda=2$, which is independent of $\alpha$, indicating a
sharp transition point. Figure \ref{fig:Lambda_N} illustrates the dependence of the critical interaction strength (\ref{Lambda_crit}) on the
particle number $N$ for the threshold value $\alpha=0.001$, $\Phi^{-1}(2 \alpha) \approx -2.8782$.

In the limit $\Lambda \rightarrow \infty$ of infinite interaction strength the approximation (\ref{zbar_app}) for $\bar{z}$ does not 
converge exactly to unity but only to a value close to one depending on $N$. This is an artifact of neglecting the fluctuations 
with $\delta \bar{z}< -1/\Lambda-1/2$ which are only relevant for very high interaction strengths. If these fluctuations are taken
into account equation (\ref{zbar_app}) becomes
\begin{equation}
\fl \qquad \bar{z}(\Lambda)  =  \left(\frac{1}{2}+\sqrt{\left(\frac{1}{2}+x^* \right)^2-\frac{1}{\Lambda^2}} \right) \left[ \Phi\left(\frac{1/2-1/\Lambda}{\sigma_N} \right)+ \Phi\left(\frac{-1/2-1/\Lambda}{\sigma_N} \right) \right]
 \label{zbar_app2}
\end{equation}
and the correct limit $\bar{z} \rightarrow 1$ for $\Lambda \rightarrow \infty$ is recovered.

\section{Summary}
\label{sec:summary}
In this paper we have described the transition to self-trapping of Bose-Einstein condensates in double wells by a single scalar quantity, 
namely the time-averaged relative population imbalance between the wells as a function of the scaled interaction strength.
Using a heuristic ansatz we have derived convenient closed form approximations for the time-averaged relative population imbalance
in the two mode Bose-Hubbard approximation that are valid in the classical (mean-field) and semiclassical parameter and time 
regimes respectively. The comparison with numerically exact results has revealed a good agreement. Our technically simple treatment
complements more rigorous numerical studies of the problem based on semiclassical phase space methods. 

\ack
The author would like to thank H. J\"urgen Korsch for valuable discussions and comments. 
Financial support as ''Boursier de l' Universit\'e Libre de Bruxelles (ULB)`` is gratefully acknowledged.

\appendix

\section{The noninteracting two-mode system}
\label{sec:app}

In the noninteracting case the dynamics of the site amplitudes is governed by the linear Schr\"odinger equations
\begin{eqnarray}
  i \hbar \dot c_\rL=- \frac{J}{2}c_\rR  + \epsilon_\rL  \, c_\rL  \label{Schrod1}\\
  i \hbar \dot c_\rR=- \frac{J}{2}c_\rL  + \epsilon_\rR  \, c_\rR \, \label{Schrod2} 
\end{eqnarray}
which are readily obtained by setting $U=0$ in the Gross-Pitaevskii equations (\ref{GPE1}), (\ref{GPE2}).
Inserting the ansatz $c_\rR(t)=\exp(\ri \Omega t)$ and eliminating $c_\rL$ yields the characteristic equation
$\hbar^2 \Omega^2+(\epsilon_\rL+\epsilon_\rR)\hbar \Omega + \epsilon_\rL \epsilon_\rR -J^2/4=0$ with the solution 
\be
   \hbar \Omega_\pm=-\frac{\epsilon_\rL+\epsilon_\rR}{2} \pm \frac{\sqrt{J^2+\Delta^2}}{2}
\ee
where $\Delta=\epsilon_\rR-\epsilon_\rL$ is the difference of the on-site chemical potentials. The amplitude in the right site can thus be written as the superposition
$c_\rR(t)=A_+ \exp(\ri \Omega_+ t)+A_-\exp(\ri \Omega_- t)$. Inserting the initial conditions $c_\rL(t=0)=1$, $c_\rR(t=0)=0$ into (\ref{Schrod1}), (\ref{Schrod2}) we 
obtain $\ri \hbar \dot c_\rR(t=0)=-J/2$. These initial conditions for $c_\rR$ and $\dot c_\rR$ lead to $A_-=-A_+$ and $A_+=J/(2\sqrt{\Delta^2+J^2})$ respectively.
Thus we arrive at 
\be
  c_\rR(t)=\frac{J}{\sqrt{J^2+\Delta^2}} \exp \left(-\ri \frac{\epsilon_\rL+\epsilon_\rR}{2 \hbar}t \right) \sin\left(\frac{\sqrt{J^2+\Delta^2}}{2 \hbar}t\right)
\ee
which implies the result (\ref{pt_lin}) for $p(t)=|c_\rR(t)|^2$.

\end{document}